\documentclass[pdflatex,sn-apa]{sn-jnl}% APA Reference Style
\usepackage{graphicx}%
\usepackage{multirow}%
\usepackage{amsmath,amssymb,amsfonts}%
\usepackage{amsthm}%
\usepackage{mathrsfs}%
\usepackage[title]{appendix}%
\usepackage{xcolor}%
\usepackage{textcomp}%
\usepackage{manyfoot}%
\usepackage{booktabs}%
\usepackage{algorithm}%
\usepackage{algorithmicx}%
\usepackage{algpseudocode}%
\usepackage{listings}%
\usepackage{tabularx}%
\usepackage{makecell}
\theoremstyle{thmstyleone}%
\theoremstyle{thmstyletwo}%

\theoremstyle{thmstylethree}%

\begin{document}

\title[Article Title]{``Can a Robot Really Care About Me?" How Perceiving a Tutor as Human or AI Affects Computer Science Students' Attitudes and Experiences}


\author{\fnm{Tymon} \sur{Vu}}\email{tvu38@calpoly.edu}
\author{\fnm{Wenxuan ``Wendy''}
\sur{Shi}}\email{wendyshi@cpp.edu}
\author{\fnm{Silas} \sur{Hsu}}\email{sihsu@calpoly.edu}


\abstract{Conversational artificial intelligence can reproduce many instructional and social behaviors of human tutors, but students may interpret those behaviors differently depending on who they believe is providing them.  We examined how perceived tutor identity shapes students' evaluations of instructional feedback, satisfaction of psychological needs according to Self-Determination Theory, and their communication strategies with tutors.  In a within-subjects Wizard-of-Oz study, 27 introductory computer science students completed two text-based tutoring sessions involving Python code-explanation problems.  Participants were told one tutor was human and the other was AI, although the same human provided feedback in both conditions.  We found participants evaluated both tutors favorably and reported few consistent differences in the instructional value of the feedback. However, participants frequently questioned whether AI could genuinely care about them, which corresponded with the human tutor having higher scores of relatedness need satisfaction.  Human presence offered connection and accountability for some participants but introduced anxiety in others.  Participants also reported communicating more casually with the human tutor and more explicitly with the AI tutor.  These findings suggest complementary roles for human and AI tutoring and highlight the importance of considering psychological needs alongside instructional quality.}




\maketitle

\section{Introduction}
Providing timely, individualized feedback to students is a persistent challenge in education, particularly in large courses where instructors have limited time to address each student's questions and difficulties. Recent advances in large language models (LLMs) have expanded the potential of AI-based tutoring systems to address this challenge. Unlike earlier systems that relied on predefined responses, LLM-based tutors can engage students in open-ended dialogue, interpret their explanations, and provide feedback tailored to their questions \citep{kasneci2023chatgpt, yan2024practical}. These capabilities are even being extended to direct instructional delivery. For example, the Alpha School uses AI-driven learning platforms to deliver academic lessons, with human supervisors only there to help navigate the AI \citep{sulkin2026}. These developments reflect growing interest in AI's potential to make personalized tutoring more accessible and scalable. 

% AI systems are increasingly used in educational settings for assessment, prediction, tutoring, student support, and learning management \citep{zawacki2019systematic, crompton2023artificial}. These systems are often motivated by practical needs: students need timely, individualized support, while instructors face constraints such as large class sizes and limited availability \citep{labadze2023role}. 
% AI-powered tutors and chatbots are being adopted to provide on-demand scalable support and personalized feedback to students in several subjects \citep{croxton2025, deng2024does}.

However, effective tutoring involves more than providing correct answers or explanations. First, tutors must respond to students' reasoning and provide appropriate feedback to help them identify and address misconceptions \citep{shute2008focus, hattie2007power}. Second, tutors often provide motivational and interpersonal support. Research on supportive teaching demonstrates that practices such as acknowledging students' perspectives and communicating confidence in their abilities can support students' motivation and engagement \citep{reeve2016autonomy, lepper2013motivational, furrer2003sense}. Finally, students themselves are active contributors to the tutoring interactions. They must articulate their understanding, explain their reasoning, and communicate their questions or confusion \citep{shute2008focus, graesser1994question}. These dimensions are interconnected: how students experience a tutors' support may influence their willingness to seek help, while the information they share can shape the guidance they receive from the tutor.

As conversational AI increasingly assumes roles traditionally performed by human educators, an important question is whether students experience these dimensions of tutoring differently depending on who they believe is providing the support. Such differences could influence not only students' perceptions of the tutor, but also their motivation and willingness to participate in the interaction. Prior research comparing human- and AI-generated feedback has examined learning outcomes and students' evaluations of feedback, with mixed results \citep{kaliisa2026how}. More directly, Nazaretsky et al. found that revealing whether educational feedback was produced by a human or AI influenced students' evaluations, particularly the perceived genuineness of AI-generated feedback \citep{nazaretsky_who_2026}. However, existing research has largely focused on feedback provided on students' work rather than ongoing, back-and-forth tutoring interactions.  Consequently, less is known about how perceived tutor identity shapes students' psychological experiences and communication during interactive tutoring. Understanding these differences could inform how AI tutors should communicate with students and how human and AI support can best complement one another.

% compare feedback from different providers or examine students' judgments of feedback presented outside an ongoing tutoring dialogue. Less is known about how perceived tutor identity shapes students' broader tutoring experiences, including their psychological need support and how they communicate with the tutor, when differences in the underlying provider are controlled.  

% Knowing these differences is key to chatbot design and deployment strategies. % Why is it important? Implications of knowing the answer to this question? Maybe chatbot designers be more informed beyond "emulate humans" , maybe we know the best situations to deploy AI vs humans...?
% Research on ChatGPT-supported learning has reported positive effects on academic performance and affective-motivational outcomes, suggesting that these systems may offer benefits beyond simply making instructional assistance more accessible \citep{deng2024does}. 

To investigate how perceived tutor identity shapes students' experiences during interactive tutoring, we address the following research questions:

% However, these deployments raise questions as to...

 % existing literature measures X Y Z. fewer work examines affective and motivational outcomes... and does not compare human vs AI directly... also the tutoring context as opposed to AI integration in the the curriculum....

 % One paragraph on related work?
\begin{itemize}
    \item RQ1: How do students perceive and evaluate the instructional feedback of human vs AI tutors?
    \item RQ2: How does feedback from human vs AI tutors satisfy students' psychological needs?
    \item RQ3: What differences in behaviors do students exhibit between human and AI tutors?
\end{itemize}

% Existing work has largely examined how AI tutors and chatbots impact students' academic achievement \citep{deng2024does}

To address these questions, we conducted a within-subjects ``Wizard of Oz'' study with 27 students enrolled in introductory computer science courses. Participants completed two text-based tutoring sessions involving Explain-in-Plain-English (EiPE) problems, which require students to describe the behavior of Python code in English.  Participants believed they were interacting with a human tutor in one session and an AI tutor in the other, although the same human tutor provided feedback in both conditions using a common tutoring protocol. After each session, participants completed questionnaires assessing perceived autonomy support, competence and relatedness need satisfaction, and enjoyment. We also conducted interviews to examine students' evaluations of the tutors and their reported communication strategies. 

Overall, participants evaluated both tutors positively and described many similarities in the instructional feedback they received. Median ratings of autonomy support, relatedness, and enjoyment were higher for the human-labeled tutor, although none of the differences were statistically significant after correction for multiple comparisons. Our interviews provided converging evidence for some of these patterns, particularly relatedness. Interview participants frequently questioned whether an AI could genuinely care about them, sometimes perceiving its praise or encouragement as artificial. The perceived presence of a human tutor also had varying effects: it could provide a sense of connection and motivate students to work harder, but it could also create pressure to avoid mistakes or asking too many questions. Finally, some participants reported communicating more casually with the human tutor while using more direct and transactional language with the AI, reflecting different expectations about each tutor's ability to understand their questions.

 % Higher feelings of relatedness for the human...
 % Similar for instructional value
 % More transactional, less ambiguous to the AI

Our findings suggest that designing supportive AI tutors requires more than reproducing the instructional and conversational behaviors of human tutors, and suggest the need for future work to further understand students' individual preferences for social cues and encouragement.  For educators, our findings suggest potentially complementary roles for human and AI tutoring: AI may provide a lower-pressure environment for seeking assistance, while human tutors may offer a stronger sense of interpersonal connection and accountability that some students find motivating.  More broadly, our study highlights the importance of evaluating AI tutors not only through perceived instructional quality, but also through students' psychological needs and participation in the learning interactions.

% What's new: a comprehensive evaluation, we draw on SDT
% A direct comparison
% investigates how the *perception* of identity.
% A tutoring context, while others have been more on general classroom context

\section{Related Work}
% https://drive.google.com/drive/folders/15tmZI3jfDkUBXRpV6lrarUUIc8B4gmfp

We first review the development of AI tutoring and feedback systems, particularly in computing education, followed by research on instructional feedback and how students engage with it. We then examine the motivational and interpersonal dimensions of tutoring, drawing on Self-Determination Theory (SDT). Finally, we review studies comparing human and AI feedback, focusing on how perceptions of the feedback provider influence students' evaluations.

\subsection{AI Tutoring and Feedback in Education}
Intelligent tutoring systems (ITSs) have long been developed to provide individualized instruction and feedback, allowing students to receive guidance without relying exclusively on instructor availability. Prior research has shown that these systems can support learning when they provide adaptive explanations, hints, and opportunities for students to correct their understanding \citep{vanlehn2011relative, ma2014intelligent, d2013autotutor}. More recently, advances in large language models (LLMs) have expanded the capabilities of educational AI systems, allowing them to respond conversationally to open-ended questions and generate feedback tailored to students' written responses \citep{kasneci2023chatgpt, yan2024practical}. These developments have renewed interest in using AI tutors to provide accessible, individualized support across educational settings \citep{crompton2023artificial, labadze2023role, okonkwo2021chatbots, shetye2024evaluation, kane2025bespoke}.

Computing education provides a particularly relevant context because students frequently need immediate feedback on code, debugging, code comprehension, and conceptual explanations. Automated grading and feedback tools have long been used to provide this support at scale within computing education \citep{keuning2019systematic}. Researchers have deployed and evaluated AI systems for autograding \citep{fowler2025evaluating}, LLM-generated programming feedback \citep{gabbay2024combining}, and AI-assisted pair programming \citep{fan2025impact}. Recent work has also explored richer forms of LLM-based programming support that encourage students to solve problems independently \citep{kazemitabaar2024codeaid, liffiton2023codehelp}. These systems illustrate the potential for conversational AI to provide individualized instructional assistance, while simultaneously raising questions about how students interpret and use the feedback they receive.

\subsection{Instructional Feedback and Student Engagement}
Educational research has extensively examined how formative feedback helps students recognize errors, improve their understanding, and make progress toward learning goals. Effective feedback may include identifying incorrect responses, explaining underlying concepts, providing hints, or suggesting strategies for improvement \citep{hattie2007power, shute2008focus}. However, feedback is increasingly understood as an interactive process rather than simply information transmitted from an instructor to a student. Research on feedback literacy emphasizes students' active role in understanding feedback, evaluating their work, managing emotional responses, and taking action to improve \citep{carless2018development}. In a critical scoping review, van der Kleij and Lipnevich identified students' feedback perceptions as an important component of feedback engagement, while finding limited and inconsistent evidence linking these perceptions to feedback use and learning outcomes \citep{vanderkleij2021student}. Students may perceive feedback as accurate, helpful, or appropriately detailed without necessarily achieving greater understanding or adopting the feedback.  This distinction is particularly relevant to AI-generated feedback, where students may also need to evaluate the reliability of the information provided. For example, Li et al. investigated how erroneous AI autograding feedback affected students working on EiPE problems, demonstrating the potential consequences of inaccurate feedback for learning and students' responses to the grading system \citep{li2023wrong}.

% Feedback that is too vague may provide insufficient guidance, while overly detailed explanations may be unnecessary or reduce opportunities for independent problem solving \citep{shute2008focus}. In conversational tutoring, these considerations are especially important because guidance must be adapted to students' changing understanding and requests for assistance.

In tutoring, students take an active role not only in interpreting feedback, but also shaping the guidance they receive. By explaining their reasoning, asking questions, and seeking clarification, students help tutors identify where support is needed. These behaviors can themselves support learning: self-explanation can deepen understanding \citep{chi1994eliciting}, while the quality of students' questions during one-on-one tutoring is associated with learning outcomes \citep{graesser1994question}. Seeking useful assistance also requires students to recognize their difficulties and decide what kind of guidance would help them to make progress. Educational psychology research distinguishes instrumental help-seeking, which aims to develop understanding and independent problem-solving, from seeking help primarily to obtain answers \citep{gall1985helpseeking, newman1994adaptive}. In interactive learning environments and ITSs where students can ask for hints on demand, students often demonstrate the latter strategy \citep{aleven2003help, marwan2020unproductive}. 

% Prior work has also shown that ITSs can support students in developing more effective help-seeking strategies \citep{aleven2016help}. 

Conversational AI introduces additional challenges into this process because students have considerable flexibility in how they formulate requests and respond to the system. Tankelevitch et al. highlight the metacognitive demands of generative AI, including articulating goals, evaluating responses, and determining how to rely on AI assistance \citep{tankelevitch2024metacognitive}. Studies of computing students using AI programming assistants have also documented variation in the types of assistance students request and how they engage with the responses \citep{yang2024debugging, sheese2024patterns}. Thus, understanding students' experiences of tutoring requires examining not only how they evaluate the feedback they receive, but also how they communicate their needs and engage with the guidance provided.

\subsection{Psychological Needs and Support}
Beyond learning performance, prior work has examined how AI can emotionally support and motivate students. For example, affective tutoring systems have been designed to respond to novice programmers' emotional states \citep{fwa2018architectural}, while AI-assisted pair programming can affect intrinsic motivation and programming anxiety \citep{fan2025impact}. A recent meta-analysis found that ChatGPT-based interventions can improve affective-motivational outcomes relative to unsupported learning conditions \citep{deng2024does}. Although these studies do not necessarily examine psychological need satisfaction directly, they suggest that evaluating AI tutoring requires considering motivational experiences alongside instructional outcomes.

Self-Determination Theory (SDT) offers a specific framework for examining motivational and affective outcomes. SDT proposes that intrinsic motivation and well-being is supported by satisfaction of three basic psychological needs: autonomy, competence, and relatedness \citep{ryan2020intrinsic}.  In educational contexts, satisfaction of these needs is associated with more autonomous forms of motivation, including intrinsic motivation, as well as greater engagement, well-being, and academic achievement  \citep{taylor2014intrinsic,kaplan2012enhancing, niemiec2019need,yu2016teacher}.

Autonomy refers to experiencing one's actions as volitional and self-endorsed.  Examples of ways teachers can support autonomy include avoiding controlling language, eliciting students' perspectives, providing meaningful rationales for course material and giving meaningful choices to students \citep{reeve2016autonomy,patall2018daily,assor2002choice,reeve2006teachers}.  Competence refers to feelings of mastery, learning, and self-improvement.  Clearly structured learning activities, appropriate scaffolding, informative feedback, and clear expectations can support students' competence \citep{ryan2020intrinsic}.  Relatedness in educational contexts refers to a sense of belonging and high-quality relationships between students and their teachers, peers, and parents \citep{furrer2003sense,hartnett2015influences}.

% Annamalai and Nasor: interviews with students who reguarly engaged with ChatGPT in their coursework
% Annamalai: Exploring English language learning via Chabot: A case study from a self determination theory perspective -- also interviews
% Chiu: A classification tool to foster self-regulated learning with generative artificial intelligence by applying self-determination theory: a case of ChatGPT. perception from expert panel of teachers that ChatGPT could support autonomy and competence, but not student needs for relatedness well.
% Ma and Chen: Comparison study. AI gamified language learning.  Teacher scaffolding did best on Competence and Autonomy, with a lesser boost to relatedness
% Schweder et al: ChatGPT supported teaching environment had higher SDT needs satisfaction on autonomy and competence, but not relatedness. Comparison study.

Recent research has begun to explore how generative AI might support psychological needs from a SDT perspective, although the evidence remains limited due to the recency of the technology.  Annamalai and colleagues' interviews of student ChatGPT users \citep{annamalai2023exploring,annamalai2025exploring} reveal that generative AI tools may support autonomy by allowing students to choose when, how, and at what pace they seek assistance, and may support competence by providing immediate explanations, feedback, and help tailored to students' question. They may also support relatedness by offering social advice and virtual companionship free of social pressure.  But at the same time, some students find it hard to engage emotionally with a chatbot.

Other work has compared instructional environments that integrate AI to control groups without AI integration \citep{yi2025human,schweder2026motivation}.  Students that experience an AI-integrated environment exhibit greater autonomy and competence need satisfaction, but relatedness gains are much less pronounced.  In contrast to these studies, we examine one-on-one tutoring, and directly compare human and AI feedback.  In addition, these studies conceptualize relatedness as overall connection between students and their teachers and peers at school, while the current study focuses measurement of relatedness on the one-on-one relationship between student and tutor.

\subsection{Human vs. Automated Feedback}

%Michael J Curran, Peter Draus, George Maruschock, and T Maier. 2013. Student perceptions of project essay grade (PEG) software.
Research comparing human and automated instructional feedback predates recent advances in generative AI. Earlier studies examined students' responses to automated feedback and whether attributing feedback to a computer rather than a human influenced their perceptions and learning.  For example, Lipnevich and Smith manipulated whether detailed feedback on students' essays was attributed to an instructor or a computer and found that the perceived source of feedback had little effect on students' performance \citep{lipnevich2009effects}. However, a follow-up study revealed that some participants who believed they got computer feedback discounted praise, believing that the “computer” was merely programmed to provide the same positive feedback to everybody \citep{lipnevich2009really}. In interactive programming environments, Price et al. identified trust, perceived tutor capabilities, accessibility, and independence as factors shaping students' help-seeking behaviors with human tutors vs. an ITS \citep{price2017factors}. These findings suggest that the perceived source of feedback can influence students' experiences in ways that are not necessarily reflected in learning outcomes.

% Simply believing that information comes from a computer rather than a person can alter cognitive performance and the achievement goals students adopt \citep{spatola2021human}.

% This body of work shows that source attribution has long mattered in educational feedback contexts: students may interpret the same feedback differently depending on whether it appears to come from a human instructor or a computer.

Recent advances in generative AI have renewed interest in comparing human and automated feedback. A recent meta-analysis comparing AI and human feedback across educational domains found no statistically significant overall differences in learning performance or feedback perceptions, but reported high variability across studies \citep{kaliisa2026how}. Individual studies likewise suggest that the relative effectiveness and perceived value of feedback depend on the context and outcomes examined. For example, Er et al. found that students receiving instructor-generated feedback in a Java programming course perceived it as more useful and achieved greater improvements in their revised assignments than those receiving GPT-4 feedback, despite being unaware of the feedback's source \cite{er2025assessing}. In contrast, other studies have reported comparable learning outcomes between human- and AI-generated feedback \cite{escalante2023ai}. Beyond learning performance, students may value different qualities of each provider. Henderson et al. found that university students generally considered both AI and teacher feedback helpful, but trusted teacher feedback substantially more \cite{henderson2025comparing}. These findings highlight the importance of distinguishing the instructional effectiveness of different feedback providers from students' perceptions of them. 

% At the same time, some studies find that AI and human feedback produce similar preferences and learning outcomes \citep{escalante2023ai} and that credibility and empathy cues can improve perceptions of AI feedback  \citep{kotlyar2025student}. These findings suggest that students' responses to AI feedback depend on the context and framing of the feedback, not only on whether the source is human or AI. 
Comparisons of human- and AI-generated feedback often involve differences in the feedback content and quality, making it difficult to determine how much students' evaluations are influenced by the provider's identity itself. To isolate the role of perceived identity, some studies have manipulated whether the same feedback is attributed to a human or AI. For example, Ruwe and Mayweg-Paus found that feedback attributed to AI was associated with greater perceived provider trustworthiness than feedback attributed to humans, particularly regarding expertise \citep{ruwe2023argumentation, ruwe2024embracing}. In contrast, Abendschein et al. found that a human-attributed grader was rated more highly on competence, trustworthiness, and goodwill \cite{ruwe2023argumentation, ruwe2024embracing, abendschein2024credibility}. However, these differences did not consistently extend to evaluations of the feedback itself.   Using a different approach, Nazaretsky et al. examined students' evaluations of actual human- and AI-generated feedback before and after disclosing its source. They found that students' ability to distinguish AI from human feedback depended on the task, and that AI-generated feedback was evaluated less favorably after disclosure, especially on genuineness, while human feedback providers were rated as more credible than AI provider \citep{nazaretsky_who_2026}. Beyond judgments of accuracy or trustworthiness, identical messages are rated as less empathic and supportive when attributed to AI than to a human, and even an unprompted suspicion that AI was involved is enough to reduce their perceived support \citep{rubin2025comparing, yin2024ai}. Thus, this body of work demonstrates that perceived source identity can influence students' perceptions of feedback, but its effects depend on the dimensions being assessed and do not consistently favor human or AI sources.

However, much of this research has focused on feedback provided on students' work rather than ongoing tutoring interactions in which students can communicate their reasoning, ask questions, and respond to guidance. Less is known about how perceived tutor identity relates to students' psychological needs and communication strategies during open-ended dialogue. Our study investigates these dimensions through a Wizard-of-Oz experiment in which the same underlying human tutor provides instructional support under perceived human and AI identities, reducing differences in instructional expertise and approach.

% These effects matter especially for educational AI because tutoring is not only informational but also relational. Although users can respond socially to computers, they do not necessarily treat AI systems as equivalent to human partners. Clark et al. found that many users described conversational agents as ``just a tool,'' suggesting that beliefs about machine agency can constrain whether users experience these systems as socially meaningful \citep{clark2019makes}. Students may want accurate feedback, but they may also want to feel understood, respected, and free to express confusion, dimensions of a tutoring interaction that an ``AI'' label may lead them to discount.

% Existing work on human versus AI feedback has therefore established that source labels can influence constructs adjacent to self-determination theory, such as credibility, trust, and perceived genuineness. Rather than asking only whether AI feedback is trusted or useful, we ask whether an AI-labeled tutor supports students' autonomy, relatedness, competence, and enjoyment to the same degree as a human-labeled tutor.

% Much of this evidence, however, rests on short-term or one-time judgments and on self-reported evaluations of the feedback, which are known to track its actual influence on learning only loosely \citep{evans2013making}. Less is known about how source labels affect the psychological needs that underlie intrinsic motivation, such as autonomy, competence, and relatedness, over the course of a sustained tutoring interaction. 

\section{Methods}
To answer our research questions, we conducted a within-subjects study in which participants interacted via text chat with tutors presented as human and AI, completed post-session questionnaires, and participated in a semi-structured interview. The interview guide, programming problem bank, and tutoring protocol are provided in Online Resource 1.

\subsection{Participant recruitment}
Participants were recruited via in-class announcements in our institution's first and second introductory programming courses intended for computer science majors (CS1 and CS2).  These courses ran on the quarter system, 10 weeks in duration, in Fall 2025 and Winter 2026.  Recruitment occurred in week 6, and the experiment ran during weeks 7–10, after the courses had introduced the relevant programming constructs that appear in our problem bank (e.g. loops and dictionaries).  In total, 27 students participated: 14 from CS1 and 13 from CS2.  Table \ref{table:demographics} summarizes the demographics.  The study lasted a median of 86 minutes (mean = 84 minutes) and each participant received a \$30 gift certificate as compensation.

\begin{table}[h]
\begin{tabular}{| m{0.20\linewidth} | m{0.75\linewidth} |}
    \hline
    \textbf{Attribute} & \textbf{Distribution} \\ \hline\hline
    Age & 19 years: 22 (81.5\%); 18 years: 5 (18.5\%) \\ \hline
    Gender & Male: 15 (55.5\%); Female: 9 (33.3\%); Nonbinary: 2; Prefer not to answer: 1 \\ \hline
    Race & Asian: 12 (44.4\%); White: 10 (37.0\%); Hispanic: 8 (29.6\%); Black: 1; Pacific Islander: 1; Don't Know: 1 \\ \hline
    Expected grade in their CS class & A: 15 (55.5\%); B: 10 (37.0\%); C: 2 (7.4\%)\\ \hline
\end{tabular}
\caption{Self-reported demographics of our 27 participants.}
\label{table:demographics}
\end{table}

\subsection{Explain in Plain English (EiPE) problems}
Participants in our experiment received tutoring in code reading and explanation skills through working on \textit{Explain in Plain English} (EiPE) problems.  EiPE problems present a short block of code and ask for a concise, high-level English description of what the code does.  We chose EiPE because it is not part of our institution's computer science curriculum, increasing the likelihood that student participants would feel challenged and receive meaningful feedback from the tutoring.

At the start of the experiment, all participants received the same introduction to EiPE problems.  This was intended to reduce the amount of basic EiPE instruction needed during the first tutoring session and, in turn, reduce ordering effects across sessions.   Participants first read a document containing a description of the structure of EiPE problems, tips on how to solve them, and mistakes to avoid.  They then completed a practice problem and received facilitator feedback.  Both the facilitator and tutor's feedback followed EiPE standards from prior work \citep{fowler2021should,li2023wrong}; responses needed to be high-level and unambiguous to be considered completely correct.  %Appendix REF! contains the practice problem used in the study and illustrates the grading standards used.

%1 out of 27 participants formulated the correct answer to the practice problem on their first try, signaling the potential benefits that our participants could receive from tutoring.

The tutoring sessions used a bank of Python EiPE problems drawn from  the EiPE literature \citep{fowler2021should,li2023wrong} and problems authored by the research team.  The bank contained 3 easy problems, 8 medium problems, and 4 hard problems.  In addition, each problem was annotated with one of the programming patterns described in Iyer et al. \citep{iyer2021pattern}, which the tutor used to inform problem selection (see Section \ref{section:problem_selection}).  The Supplementary Materials contain all problems.

%We classified difficulty holistically using several heuristics: (1) indentation depth, which prior work associates with difficulty [citation]; (2) how frequently the problem's programming pattern appears in introductory courses \citep{iyer2021pattern}; and (3) how late the relevant programming constructs appear in our institution’s curriculum, with later-introduced constructs such as dictionaries treated as more difficult (4) experiences from pilot experiments.

\subsection{Tutoring sessions}

Participants were told that they would work with both a human tutor and a conversational AI tutor, with the order counterbalanced across participants.  All tutoring happened via text chat in Discord.  As a cover story, we expressed that we were testing a new, custom AI tutor.  In addition, to explain the relatively long delay that participants would experience in the AI condition, we told participants ``we’ve purposefully delayed the AI's response time to give you more time to reflect on your answers.''

A single researcher assumed both the human and AI tutor personas.  The researcher had served as the lead instructor for college courses, had experience grading EiPE responses, and was familiar with motivationally-supportive tutoring \citep{lepper2013motivational} and autonomy-supportive teaching practices \citep{reeve2016autonomy,reeve2006teachers}.

The tutor followed a semi-structured tutoring protocol (excerpt in Table \ref{table:tutor_responses}) designed both to promote consistency across conditions and to incorporate literature-informed practices.  The protocol included multiple scripted sentence starters for common scenarios, such as greetings, goodbyes, incorrect reasoning, requests for line-by-line explanations, ambiguous answers, and others.  It specified a progression from minimally informative hints to more explicit guidance, preserving opportunities for students to solve problems independently and therefore autonomy.  When students solved a problem without hints, the tutor reinforced their agency (e.g. ``Nice work, you did that without help!'').  After each problem, the tutor gave students a choice of difficulty over the next problem, or suggested a difficulty in an open manner (e.g. ``Am I right to guess you're looking for a harder problem next?'').

This protocol served as a set of conversational anchors and semantic/pedagogical guidelines rather than a strict script.  Although recurring elements such as greetings and goodbyes generally followed the provided language, the nature of tutoring routinely required improvisation, especially in situations the protocol did not cover or to ensure natural conversational flow.  For example, the tutor often asked students to reason through a simplified version of a problem before returning to the original, as question-specific hints were not included in the protocol.  Regardless, the tutor saught to use their instructional judgment to preserve the protocol's motivational and autonomy-supportive principles across both conditions.

\begin{table}[h]
\begin{tabular}{| p{0.20\linewidth} | p{0.75\linewidth} |}
\hline
\textbf{Situation} & \textbf{Response options} \\
\hline\hline
Greetings & * Hello! How is your day going? \newline * Hello! How are you doing today \\
\hline
Correct answer & * That looks good to me! Nice work. \newline * That seems right. Good job. \newline * (If they struggled) Great job, that was a particularly tricky one.  Good for you! \\
\hline
Correct answer on the first try & * Nice work, you did that without help! \newline * Nicely done, and without help! Good for you! \\
\hline
Correct answer follow-ups & * Here are other ways you could describe the code: \newline
* Here are some example correct answers I have on record: \newline
* (After revealing answers) How are you feeling after that? \newline
* (After revealing answers) Do you have any thoughts before we move on? \newline
* (After revealing answers) How much would you agree with my assessment? \\
\hline
Ambiguous wording (initial) & * I think I know what you mean, but the wording of the response still has issues. What do you think the issue might be? \newline * This is close! But there’s still an issue with the phrasing. Any guess what part might need adjusting? \newline * This is on the right track!  But if someone tried to deliberately misinterpret your description, how might they do it? \newline * (Six other starters omitted) \\
\hline
Ambiguous wording (hint) & * I can see a gap between what you mean and what the words literally say. (Insert details) \newline * I think more specificity could improve your answer. (Insert details) \newline * (Three other starters omitted) \\
\hline
\end{tabular}

\caption{Excerpt from the tutor's script.  Multiple alternative responses were provided for each situation to help prevent individual participants from experiencing repetition.}
\label{table:tutor_responses}

\end{table}

Initially, the tutor used separate user accounts for each condition (one named ``AI Tutor'' and the other ``Human Tutor'') and greeted students acting as the corresponding persona.  However, after the 10th participant, the researcher acting as the tutor raised concerns that their awareness of the experimental condition could bias tutoring behavior.  Thus, beginning from the 11th participant, both conditions used the same account labeled ``Tutor'' and a more generic set of greetings, which blinded the tutor to the assigned condition.

As a check, we qualitatively coded the tutor's language in the chat logs using a deductive strategy.  Codes came from the supportive tutoring strategies described by Lepper et al (e.g. ``control over help'', ``emphasize partial success'', ``emphasize the difficulty of the task'') \citep{lepper2013motivational}, and the use of informal and emotional language inspired by Clark et al (i.e. ``humor'', ``emojis'', ``shared experiences'') \citep{clark2019makes}.  A chi-square test of frequency of supportive behavior and informal/emotional language before and after the change found no significant differences ($p=0.787$).  In addition, we calculated the differences in number of tutor messages, words per tutor message, and session duration between the human and AI conditions.  Welch's t-tests found no clear evidence of differences in these metrics before vs after tutor blinding ($p=0.15$, $p=0.71$, and $p=0.13$ respectively).

\subsubsection{Problem selection and session duration}
\label{section:problem_selection}
The tutor randomly selected a medium-difficulty problem to start each session.  After this initial problem, student choice guided subsequent problem selection.  To support autonomy, the tutor always asked which difficulty level the student wanted next, or suggested a higher level and asked for confirmation.  The only restriction in the first session was that the tutor avoided at least one programming pattern among those present in the medium-difficulty pool.  This reserved pattern was then used for the initial medium-difficulty problem in the second session to reduce cross-session learning effects.

Each tutoring session targeted a 30-minute duration. If a participant passed the 20-minute mark and the tutor judged that starting another problem would substantially exceed the target duration, the tutor ended the session after the participant completed their current problem. Once a session reached 30 minutes, the tutor told the student that time was running short and offered more substantial hints to help the participant finish the problem.  The tutor would also end the session after four completed problems, no matter the elapsed time, to avoid exhausting the problem bank.

\subsection{Outcome measures and debrief}
After each tutoring session, participants completed an experience questionnaire about the tutor with whom they just worked.  All survey items can be found in Appendix \ref{appendix:survey_items}.  The questionnaire measured experiences related to SDT's three psychological needs: autonomy, relatedness, and competence.  Participants' perceptions of tutor autonomy support was measured using the 6-item version of the Learning Climate Questionnaire (LCQ), with all instances of the word ``instructor'' changed to ``tutor.''

Competence and relatedness items were adapted from Fedesco’s adaptation of the Basic Psychological Need Satisfaction Scale (BPNSS) to the educational setting \citep{fedesco2019connections}.  To adapt the Relatedness scale, we dropped items referring to the respondent's student peers, leaving only the questions about the respondent's relationship with the instructor.  We also dropped ``I am not close to the instructors in this course'' because participants were unlikely to have time to develop a close relationship with the tutor during a single session.  Lastly, we replaced all instances of ``instructor'' with ``tutor''.  %The final Relatedness scale consisted of 4 items ($\alpha = $ [placeholder]).

To adapt the Competence scale, we replaced instances of ``in this course'' with ``in this session'' and dropped ``People in this course tell me I am good at what I do'' because the experiment did not include this social evaluative component.% The final Competence scale consisted of 2 items (Spearman-Brown reliability = [placeholder]).

Lastly, besides the SDT-related constructs, we assessed overall enjoyment using 7-point Likert agreement with the statements of ``I was satisfied with the tutor'' and ``I enjoyed working with the tutor.''

After all tutoring sessions and their associated questionnaires concluded, the facilitator conducted a semi-structured interview.  Topics included general impressions of each tutor, explanations of differences in questionnaire responses between the human and AI tutors, and strategies the participant used to communicate with each tutor.  Participants used their chat log as an artifact during the interview.

After the interview, participants completed a demographics survey collecting age, race, and expected grade in their computer science course.  Lastly, the facilitator conducted a debrief that explained the study's deceptive components and gave participants the opportunity to withdraw their data; none chose to do so.

\subsection{Qualitative data analysis}

Interviews were transcribed, then analyzed using an iterative, inductive approach using Atlas.TI qualitative coding software.  One researcher first read a sample of transcripts from early, midway, and late in data collection for familiarization, and assigned descriptive codes to passages relevant to our RQs.  As additional transcripts were examined, codes were added, merged, and refined, and previously coded transcripts were revisited to apply the evolving definitions consistently.  Finally, codes were classified into higher-level themes.

The same researcher who acted as the tutor conducted the qualitative analysis.  Their experience, as well as access to the original chat logs, provided contextual knowledge for interpreting participants' comments.  The developing codebook was also shared with the researcher who conducted the interviews, who remained available to clarify interview context.

\section{Quantitative Results}
\begin{figure}[t]
        \centering
        \includegraphics[width=0.6\textwidth]{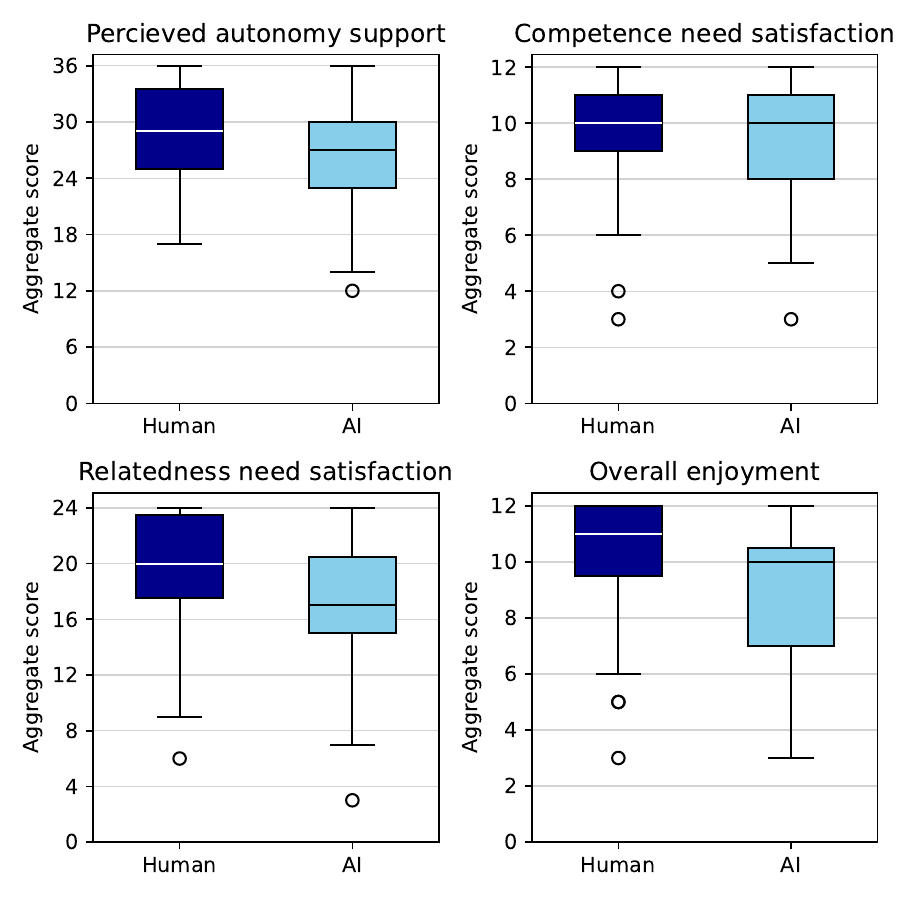}
        \caption{Aggregate scores for each of our four survey constructs, by tutor identity.  The y-axis ranges from zero to the maximum possible score for each construct.}
        \label{fig:quant_results}
   \end{figure}

    Participants evaluated both tutoring conditions favorably on the four survey constructs (Figure \ref{fig:quant_results}).  Responses were coded as integers from 0 to 6 and summed within each construct.  When expressed on a 0 - 6 response metric, median composite scores across the four constructs ranged from 5 to 6.
    
    Table \ref{table:quant_data} presents detailed statistics.  We observed higher median ratings in the human-tutor condition for perceived autonomy support, relatedness need satisfaction, and overall enjoyment; median competence need satisfaction was equal between conditions.  We used Wilcoxon signed-rank tests to compare the conditions on each of the four constructs.  None were statistically significant at a level of $\alpha=0.05$ under Bonferroni correction (critical $\alpha=.0125$).
    
   \begin{table*}[h]
    \centering
    \caption{Aggregate statistics between human and AI conditions. p-values and effect sizes computed from Wilcoxon signed-rank tests.}
    \label{table:quant_data}
    \begin{tabularx}{\textwidth}{|X|c|c|c|c|c|c|}
    \hline
    \textbf{Construct} & \textbf{Max score} & \textbf{Mdn (Hum.)} & \textbf{Mdn (AI)} & \textbf{p-value} & \textbf{Effect ($r$)} \\ \hline
    Perceived Autonomy Support & 36 & 29.0 & 27.0 & 0.0787 & 0.34 \\ \hline
    Competence Need Satisfaction & 12 & 10.0  & 10.0  & 0.8048 & 0.05 \\ \hline
    Relatedness Need Satisfaction & 24 & 19.0 & 16.0 & 0.0380 & 0.40 \\ \hline
    Overall Enjoyment & 12 & 11.0  & 10.0 & 0.0482 & 0.38 \\ \hline
    \end{tabularx}
    \end{table*}

\begin{figure}[h]
        \centering
        \includegraphics[width=0.8\textwidth]{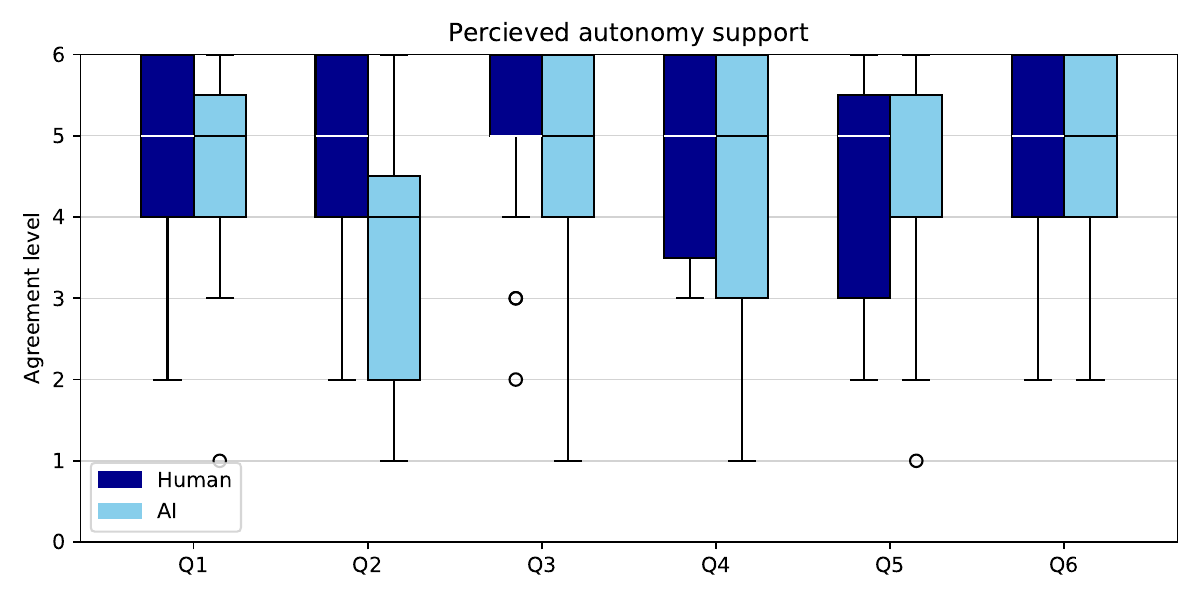}
        \caption{Distribution of perceived autonomy support scores by survey item and tutor: Q1 - I feel that the tutor provided me choices and options; Q2 - I feel understood by the tutor; Q3 - The tutor conveyed confidence in my ability to do well; Q4 - The tutor encouraged me to ask questions; Q5 - The tutor listened to how I would like to do things; Q6 - The tutor tried to understand how I saw things before suggesting a new way to do things.}
        \label{fig:autonomy_breakdown}
    \end{figure}

\begin{figure}[h]
        \centering
        \includegraphics[width=0.8\textwidth]{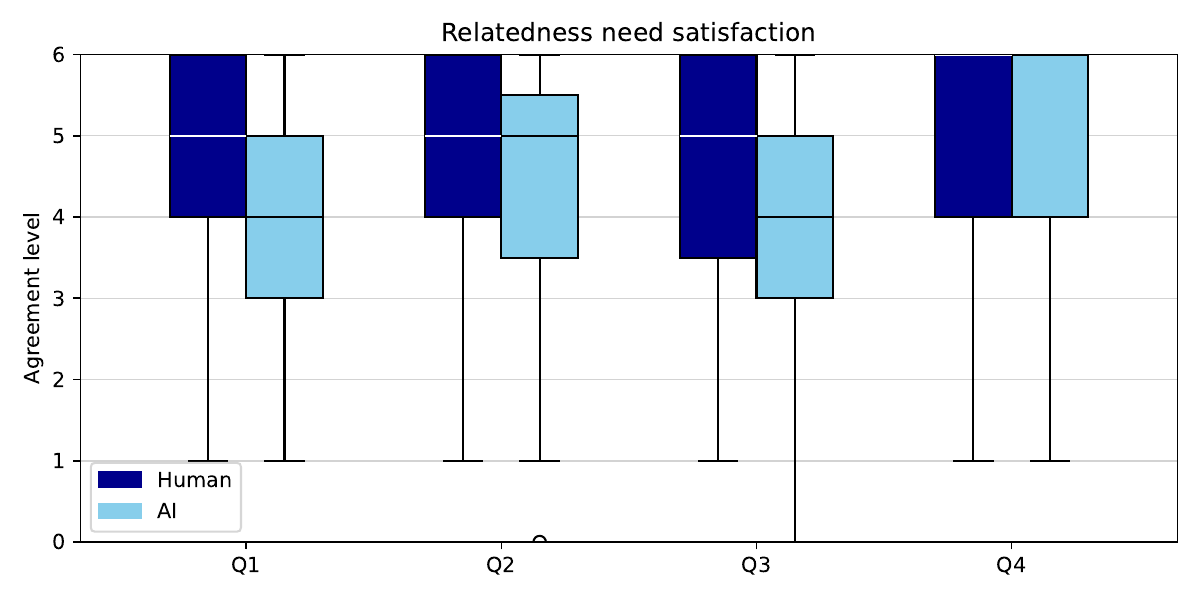}
        \caption{Distribution of relatedness need satisfaction scores by survey item and tutor: Q1 - I really liked the tutor; Q2 - I got along with the tutor; Q3 - The tutor cares about me; Q4 - The tutor does not seem to like me much (reverse coded)}
        \label{fig:relatedness_breakdown}
    \end{figure}

    We further examined whether certain survey items were more responsible for differences than other items (Figures \ref{fig:autonomy_breakdown} and \ref{fig:relatedness_breakdown}).  For autonomy, the most pronounced difference appears in the statement of ``I feel understood by the tutor.''  For relatedness, the most pronounced difference appears in the statements of ``I really liked the tutor'' and ``The tutor cares about me.''
    
\section{Qualitative Results}
%One interview (P14) was cut short
This section presents the major themes identified in the interviews. We begin with participants' overall impressions of the two tutors.  We then examine how features of the tutoring experiences affected participants' evaluations, including the content of the feedback, tutors' mistakes, and sense of relatedness.  Finally, we consider how participants adapted their messages and behavior when interacting with each tutor.

\subsection{Participants' overall impressions and comparisons}
Consistent with the generally favorable survey ratings, 19 participants expressed a positive overall impression of the human tutor, and 20 expressed a positive overall impression of the AI tutor.  13 participants remarked on how the delivery, content, and style of the feedback felt similar.  P27 stated, ``if you didn't tell me which is which, I don't think I'd be able to tell.''  Several participants associated the human tutor's enthusiastic tone and proper capitalization and punctuation with AI communication.  For example, P08 described the human tutor as ``scripted'' while P1 said ``I felt like either the human was really trying to emulate the AI or it was AI.''

At the same time, nine participants cautioned that their perceived differences between the sessions could reflect the problems they encountered rather than the tutors alone. When explaining why they had rated the human tutor substantially higher, P12 responded, ``problem set probably'', noting that they became stuck on one AI tutor problem while the human tutor session proceeded more smoothly.  P17 likewise explained, ``it's hard to compare because I made different kinds of mistakes on each of them.''  Participants' evaluations therefore reflected the complete tutoring interaction, including the problems presented and the difficulties they experienced, rather than tutor identity in isolation.  Nonetheless, some participants did attribute differences to tutor identity, which we discuss in the following sections.

% We could discuss the time it took, but it's not really relevant to our RQs.
% Not enough info, and problem variation

\subsection{Getting to the right answer and pedagogical value}
% Covers appreciates efficient feedback and instance of feedback helpful
% second part covers inefficient feedback and appreciates detail in feedback

Across both tutors, many participants expressed the feedback was overall ``efficient'', ``concise'', or ``straightforward'' at helping them move towards a correct answer.  They often pointed out instances of useful feedback and hints as they reviewed their chat logs.  Patterns of tutor behavior that participants described as pedagogically successful included checking in after the student had been idle for several minutes of inactivity (1 participant commenting on the human tutor, 2 on the AI), follow-up prompts about a problem after the answer was revealed (1 participant commenting on the human, 2 on the AI, 2 on both), and asking about the desired difficulty of the next problem (1 human, 1 AI, 3 both).  Participants especially valued feedback that built off of their answers or way of thinking, often describing such feedback as ``personalized'' (7 participants commenting on the human tutor, 7 on the AI).  Instances where the tutor recognized key conceptual errors impressed students:
% follow up prompts: "Both of them asked if I had any questions about how my answer is different [from the exemplars] or something similar to that." -P11

\begin{quote}
    \textit{There was a question where it returned a value in a loop and I got the question wrong. And somehow [the AI] pointed out and asked if I knew what happened when it returned in a loop. And I was impressed with that} -P27
\end{quote}

Interestingly, three participants perceived a greater degree of personalization from the human tutor, while no participants made the opposite comparison.  P23 felt the AI tutor ``was trying to follow a set guidelines that it was already wanting to do''.  P19 said ``this is just like a chat bot, so I felt like a lot of the responses were the same.''  To be clear, these three participants evaluated both tutors as effective; they just evaluated the human as more personalized and therefore better.
% However, we did not notice any obvious differences in the chat logs of these three participants; the tutors both built off students' answers.

%follow-ups = double-checking their understanding or asking questions about why an answer is incorrect

Other parts of the tutors' feedback style received mixed evaluations.  One hallmark of the tutors was breaking down the problem into subproblems.  Some participants (5 commenting on the human tutor, 4 on AI) perceived the feedback as detailed and aiding deep understanding.  P15 said, ``having the AI chatbot break it down really, really specifically and taking the time or looking for many ways to make me be able to understand what the code is trying to say, is what really makes me really appreciative of the tutoring.''  Other participants (1 commenting on the human tutor, 3 on the AI) felt this strategy provided unnecessary detail or went off topic, such as P16 who said ``[the human is] very much dancing around a lot of topics.''
% Both of them asked if I had any questions about how my answer is different [from the exemplars] or something similar to that. -P11

Another hallmark of the tutors' behavior was that detailed feedback arrived only after initial, more generic hints, as directed by the tutoring script.  Some participants (3 commenting on the human, 8 on the AI) appreciated the ability to solve problems on their own:

\begin{quote}
    \textit{They [human tutor] were like, ``Hey, pretty good answer, but there's something missing, and see if you can figure it out.'' And I think I kind of liked that guidance and just encouraging me to look back on it. And I feel like it makes me feel a little more accomplished.``} -P10
\end{quote}

But other participants (3 commenting on the human, 4 on the AI) described the feedback as repetitive, vague, or lacking actionable detail, or reported difficulty in interpreting the tutor's feedback.
% One student (I think 14) couldn't even answer basic questions like how many times a loop will run.

\begin{quote}
    \textit{[The AI tutor] just asked ... ``If someone were to deliberately misinterpret your description,'' which kind of didn't really help me much until they like expanded on [it].} -P6
\end{quote}

%Other than vagueness, N participants reported confusion from apparent contradictions in feedback between the two tutors.  Not including because it's not often and doesn't really contribute to the discussion.

\subsection{Reactions to tutor mistakes and grading standards}
\label{section:reactions_to_mistakes}
14 participants made comments related to real or perceived inaccuracies in tutor feedback, how the tutors responded to student questions, or the enforcement of wording standards.  Many participants got feedback that their answer was not completely correct due to ambiguities in wording.  While some participants voiced understanding or endorsement towards these standards, others complained about the tutor enforcing them: two participants complained about the AI tutor, one about the human tutor, and two about both.  These participants described the tutor as ``nitpicky'', ``a little bit arbitrary'', or as P8 put it, ``I understood what the code was doing, and the human was just like asking for a more elegant answer, which felt like... ok it's kind of a waste of time.''

Other sources of mistakes included the tutor giving a problem that was too easy after the student asked for a challenging one (3 instances), evaluating a correct answer as incorrect (4 instances), and misunderstanding students' questions or answering them incorrectly (3 instances).  Participants reported these kinds of mistakes with equal frequency for both tutoring conditions (5 instances human, 5 instances AI).  Perceiving that a tutor made a mistake lowered evaluations towards that tutor, but participants usually explained away or downplayed such instances, especially if the tutor acknowledged and corrected the mistake.

We observed no consistent difference between conditions in participants' reactions to tutor mistakes.  In two instances, however, participants explained mistakes through the tutor's perceived identity.  P22's AI tutor assessed ``creates a copy of list l disposing non-string variables'' as conceptually correct, when the correct answer should have expressed ``disposing string variables.''  P22 realized the mistake after after the tutor provided exemplar answers, and commented that the mistake was ``expected'' and ``it forgot the [correct] answer ... just AI doesn't remember a lot, I guess.''  In the second instance, the human tutor assessed P17's truly correct answer as incorrect, before quickly realizing it on their own and reversing themselves.  P17 described this incident as ``It's okay. We humans make mistakes.''  These cases suggest that participants could arrive at similar tolerance through different explanations.

\iffalse
In another two instances where the tutor did not answer questions satisfactorily, one with human and one with AI, participants blamed themselves for not being clear enough:

\begin{quote}
    \textit{I also worded [my question] funky... I think if I had pushed [the AI tutor] a little more, it would have definitely gotten there. I just didn't push a lot.} -P18 \\
    \textit{I don't know how to ask this question. So it's more on my end. I didn't know how to word my confusion.} -P3
\end{quote}
\fi

%Lastly, the tutor sometimes acknowledged or corrected mistakes, which students appreciated.

%Problem difficulty : 3 instances (2, 3, 19) // 2 human 1 AI
%Didn't evaluate a correct answer as correct: 3 instances (9, 11, 22)  // 2 human
%Tutor didn't understand a question or answered it wrong: 2 instances (3, 5, 18) -- shared responsibility code and percieved inaccurate feedback // 

%misunderstood feedback?

\subsection{Sense of relatedness to the tutors and feelings of motivation}
%Friendliness and humor
%Motivating tone
%Unmotivating tone

% Three major things: motivating and unmotivating, 

%ight participants, 2 commenting on the AI tutor and 6 on the human tutor, described the overall social interactions as nice, friendly, polite, lighthearted, or otherwise positive.  Others commented on both tutors' effective motivational support (4 participants commenting on the human, 2 on the AI, and 2 on both).  P10 described "they were very encouraging, and they offered to give me harder questions" and pointed out "Hey, I feel like you could do it."

Another theme from the interviews related not to getting the right answer or learning, but authentic social connection to the tutors and motivation from the tone of feedback.  For many participants, the interactions with the human tutor carried greater social meaning.  Eleven participants explicitly questioned whether an AI could provide genuine connection or care.  P09 explained that they got along better with the human tutor ``simply because the AI isn't real'', while P21 preferred knowing that a person ``has got my back'' over a bot ``simulating'' the same support.  P08 asked, ``can a robot really care about me?''

Perceived authenticity also shaped participants' reactions to social cues, including greetings, praise, humor, and emojis.  These cues sometimes made the AI more personable, but participants could also describe them as unexpected, artificial, or amusing.  For instance, P17 described their initial reaction to the AI's tutor greeting as ``let's just get onto the problem'', but ``actually engaged'' with the human's greeting.  P03 questioned the AI's use of ``haha'': ``I know you’re not actually laughing, so why are you telling me you’re laughing?''  P19 expressed uncertainty whether they liked the AI's feedback of ``really really close.''  They explained that it was a ``thing someone would say in real life. But I've never seen like a AI bot use it before.''

%Human feedback could also feel artificial, especially when it resembled stereotypical AI communication.  Several participants described the human tutor's polished grammar and structured responses as "robotic" or "scripted."  P1 commented \textit{I felt like either the human was really trying to emulate the AI or it was AI.}  P8's said \textit{I felt the human tutor was trying a lot to sound like ChatGPT... the answer sounded pretty scripted.}

Despite this, AI feedback provided effective motivational support to some participants.  P20 valued affirmations such as ``nice thinking'' because they communicated that the student was on the right track and provided ``more motivation to keep going.''  P07 similarly said that the AI's compliments made them feel that they could succeed despite initially lacking confidence. P02 reported taking encouragement positively ``regardless of the tutor.''

% Code: motivating (human) but excluded P16
At the same time, 4 participants reported that the human tutor provided more motivation than the AI tutor.  P15 enjoyed working with the human tutor more because ``the human connection thing'' and ``encouragement is what someone really needs to do better.''  P23 reported perceiving more emotion and complements from the human, while describing the AI tutor's complements as fewer and ``wasn't too much to motivate me further.''

Related to motivation, 5 participants reported feeling more pressure or anxiety while working with the human tutor.  P24 said ``like with an AI, I feel like I could make so many mistakes and it wouldn't judge.''  P2 expressed they did not want to ``overwhelm the human tutor with a barrage of questions'', and felt more comfortable with sending multiple messages to the AI simultaneously.  Although uncomfortable, social pressure motivated some participants.  P12 said, ``just knowing that there was a human behind it, I feel more pressure to like try harder.''  P7 noted ``I feel like knowing that it's a human tutor gave me more pressure to kind of do it right.''

Lastly, three participants (2 commenting on the AI, 1 on the human) reported that the tutor's offer of an answer was demotivating because it communicated a lack of confidence.  Both tutors started to make these offers when sessions extended past the 30 minute target.  P20 explained that one such offer ``made me feel like I wasn’t close at all'' and ``should just give up, essentially.''  P15 described an offer as sounding less like encouragement to continue and more like ``I have it if you're ready to give up.''

%One aspect included greetings, with which both tutors always started their session ("How are you doing today?").  Among participants mentioning the human tutor's greeting, 5 participants had a positive reaction and 2 had an apathetic reaction.  But among participants mentioning the AI tutor's greeting, 1 participant had a positive reaction and 2 had an apathetic reaction.

\subsection{Communication strategies and participants' conduct with the tutors}
Near the end of the interview, the facilitator probed participants about whether they observed any differences in how they conducted themselves with the human vs AI tutor.  14 participants reported no differences, but many others did.
%Participants changed how they conducted themselves depending on whether they believed they were interacting with a human or an AI. Differences appeared in their conversational style, willingness to socialize, effort, and concern about how the tutor might evaluate them.

Nine participants observed more social communication with the human tutor.  This included more colloquial and friendly language, and treating the interaction more like a conversation.  P9 explained that with the human tutor, ``there’s more of a broader range I can use to express myself instead of just like input, output, input, output.''  P11 used words such as ``awesome,'' ``sweet,'' and ``cool'' with the human tutor.  P23 felt ``a lot more comfortable just being a lot more ... casual with the human tutor.''  P1 opened the human tutor conversation with ``I'm ready for my first problem o captain my captain.''  The tutor, not anticipating this interaction, stuck with the script and failed to reciprocate this social advance.  P1 found this lack of reciprocity frustrating, expressing ``when I think [the human]'s AI, it's less frustrating to think about.''  In other words, the participant had a stronger expectation of social communication with the human.

In contrast, eight participants approached conversation with the AI with the mindset of direct and specific requests.  P15 said ``because it's AI, you would have to be more direct, more straightforward, and very specific.''  P26 ``tended to keep [questions] shorter just to make sure that it was very clear what I was asking.''

Participants also reported elaborating to different degrees depending on whether they were talking to a human or AI.  Seven participants expected that the human would be able to better infer their needs and hold more nuanced context than the AI, resulting in less explicit communication with the human and more explicit prompts to the AI.  P7 gave the AI ``more details so I get a better answer out of it,'' and believed the human would be less likely to ``mess up'' when faced with fewer details.  P26 felt that the human could ``catch nuance'' and context better than the AI.  P9 tried to make sure they spelled words correctly ``just to make it easier for [the AI] to digest my response.''  At the same time, three participants took the opposite strategy.  For instance, P16 felt the ``need to organize my thoughts before I just blurt them out [to the human]'', but expected the AI to ``organize my thoughts for me.''

\section{Discussion}
% Do as AI say: susceptibility in deployment of clinical decision-aids (Nature Article)

% Sense of Relatedness as a Factor in Children’s Academic Engagement and Performance (Furrer and Skinner)
%Citing work from attachment theory, social support, school climate, and parenting, researchers highlight the importance of caring and closeness in student–teacher relationships...
% Neither too much nor too little given away

Our findings suggest that students' experiences and behavior with tutoring depend not only on the feedback they receive, but also on who they believe is providing it.  We now discuss which of these differences are clearest and their implications for future work and design.

\subsection{Few differences in how students evaluated instructional feedback}
    We found little evidence that the identity of the tutor strongly determined how participants evaluated the helpfulness or reliability of the instructional feedback in our experimental setting.  Many participants described the tutors' feedback as highly similar, and both tutors received positive and negative comments concerning their teaching styles, mistakes, and grading standards.

    One potential area of difference that we cannot rule out, however, is reactions to tutor mistakes.  One of our participants explained an AI grading mistake in terms of limitations of AI memory, while another participant explained a grading mistake in terms of ordinary human fallibility (see end of Section \ref{section:reactions_to_mistakes}).  This suggests that students might systematically attribute human and AI tutoring errors differently, which then may affect trust and attitudes towards tutors.  Prior work in automation reliance already provides some clues on how explanations for errors might affect trust.  For instance, \cite{dzindolet2003trust} found that people quickly lose trust in an automated decision aid after seeing it err, but explaining why errors happen increases trust and reliance.  Future work can more systematically investigate errors in the tutoring context, and how they differentially affect students' evaluations.
    
    %Nevertheless, prior research outside of AI tutoring has shown that people quickly lose trust in automation after seeing it err \citep{algoAversion}, and AI grading accuracy can be perceived as worse than it actually is \citep{hsu2021attitudes}.  Future work can more systematically investigate errors under human and AI identities, and how they differentially affect students' evaluations.
    
    %One potentially important area of difference was that participants explained tutor errors through different lenses.

    % Mental model of AI affecting how much they think the advice /feedback is "personalized." or "specific to my answer"??? ASSUMING personalization is a desired quality, future work should explore whether giving students a better mental model of how the feedback is generated, which would indicate the level of personalization, increases feelings of personalization and the evaluation  -- could not find lit about percieved personalization

    %One difference we did notice involved how participants explained the cause of errors. ....  Participants tolerated errors, but the  attributed errors to different causes: broader limitations of AI in one case and ordinary human fallibility in the other.  Algorithm aversion?  Future work could test more systematically whether students attribute equivalent AI and human tutor errors to different causes and how those attributions affect subsequent verification or reliance.

\subsection{Perceived tutor identity and psychological need satisfaction}
%Competence refers to feelings of mastery, learning, and self-improvement
    Participants reported higher autonomy support, relatedness need satisfaction, and enjoyment with the human tutor, while reporting similar competence need satisfaction across conditions.  Although the differences are not statistically significant after multiple hypothesis testing correction, the interviews demonstrated that perceived tutor identity meaningfully shaped many participants' interpretation of the tutor behaviors.  Taken together, the findings suggest the existence of differences in our experimental context, especially for relatedness, but more data would be needed to estimate the magnitude of the differences accurately.
    
    The clearest convergence between the survey and interview findings concerned relatedness with the tutor.  Much of the difference in relatedness ratings arose from the statement \textit{``The tutor cares about me.''}  Correspondingly, interview participants frequently questioned whether an AI could genuinely care about them.  In addition, participants attributed greater social meaning to encouragement, humor, and conversational language when they believed these behaviors came from a human.  These findings echo prior research showing that identifying feedback as coming from AI can reduce perceived empathy and support \citep{yin2024ai, rubin2025comparing}, qualitative work showing that students have trouble treating social interactions with ChatGPT as genuine \citep{annamalai2023exploring,annamalai2025exploring}, and work suggesting that people treat conversational agents as mere tools \citep{clark2019makes}.  Together, this evidence suggests that socially supportive language alone is insufficient to produce the same sense of relatedness as interaction with a human tutor.
    
    %Prior educational research has found that perceived social threats can contribute to avoidance of academic help-seeking \citep{ryan1997should}.  In contrast, these participants described the AI tutor as less judgmental and easier to approach.  Similar findings have been observed outside education: Lucas et al. found that participants were more willing to disclose information to a virtual interviewer when they believed it was controlled by a computer rather than a human, partly because the computer framing reduced fears of negative evaluation \citep{lucas2014only}. Together, these findings suggest that human interaction can provide interpersonal connection and accountability while also introducing social pressures, whereas perceived AI interaction may reduce some of those pressures.  TODO related work on SDT.

    Perceptions of autonomy support also favored the human tutor, but with a smaller effect size; the interviews also provided less evidence of differences.  Few of our interview themes directly related to autonomy -- the most salient examples include appreciation towards opportunities to solve problems independently and control over problem difficulty.  These themes did not exhibit clear differences between the human and AI conditions.

    Differences in perceived autonomy support ratings were concentrated partially in the statement \textit{``I feel understood by the tutor.''}  Within autonomy-supportive teaching, understanding students is associated with perspective-taking and adapting instruction to their needs \citep{reeve2016autonomy}.  One hypothesis is that participants interpreted this statement in terms of interpersonal or emotional understanding, and believed that AI would be less capable of such.  This interpretation would be consistent with our findings surrounding relatedness.  Future work could analyze how students feel AI ``understands'' them, and further break this concept down into perceived understanding towards their reasoning/current knowledge, emotions, and preferences.

    Lastly, we observed equal medians of competence need satisfaction scores.  This pattern matched the interviews, in which we found few systematic differences in how participants evaluated the tutors' instructional feedback.  It suggests that conversational AI may directly emulate the content of human hints (e.g. using the strategy of breaking a problem down into subproblems) while preserving students' competence need satisfaction.

\subsection{Potential Outcomes of Different Communication Strategies with Tutors}
    While many participants reported no differences in how they communicated with the tutors, others reported more colloquial language with the human, or differences in degrees of elaboration.  While there is some related work addressing how differences in mental models and expectations can affect human communication with AI \citep{pataranutaporn2023influencing}, we must first address downstream learning outcomes to determine how best to intervene.  We hypothesize  communication strategies with AI influences learning in at least two different ways: via the information available to the tutor and via the cognitive process of constructing questions.
    %[Say more about how this connects to mental models] This suggests that students hold different mental models of the capabilities of human and AI tutors.  \cite{pataranutaporn2023influencing} similarly found that experimentally influencing beliefs about an AI's motives [ more specific, esp connection to mental models] changed how people experienced and interacted with the AI.  

    First, the amount of elaboration changes the information available to the tutor.  A student who omits relevant reasoning or context may prevent the tutor from identifying the underlying misconception, whereas a more explicit message may elicit more targeted feedback.  Modeling this relationship is challenging because AIs' responses vary with its capabilities and system prompt.  Nonetheless, future work may analyze students' prompts in terms of ambiguity and what hints a tutor could possibly provide, which would provide a more durable picture of potential effects on feedback quality.

    Second, constructing a detailed question may itself create an opportunity for learning.  Participants in our study described either elaborating their reasoning and providing context or relying on the tutor to organize incomplete thoughts and infer their intended meaning.  Prior work has found that articulating one's reasoning and formulating questions can themselves support learning \citep{chi1994eliciting, graesser1994question}.  This suggests that during the deployment of AI tutors, educators should discourage students' tendencies to routinely offload this cognitive work, especially as AI becomes better at interpreting incomplete questions and diagnosing students' confusion.

\subsection{Additional Implications for the Design and Deployment of Conversational AI Tutors}
    Besides the implications we have already mentioned, our findings suggest several additional directions for the design and deployment of AI tutoring systems.  First, educators should treat human and AI tutoring as complementary forms of support.  Some of our participants felt more motivated to work harder because another person was helping them, while other participants described the AI tutor as less judgmental and easier to approach.  Social factors such as fears of appearing incompetent and expectations of independence have long been recognized as factors that contribute to avoidance of help-seeking \citep{butler1998determinants}.  Therefore, AI tutors offer promise as an accessible, lower-pressure option for students seeking immediate instructional help.  Meanwhile, human tutors may especially benefit students whose relatedness needs are not being met or who benefit from the motivating social presence of another person.  Where human tutor capacity is limited, educators could provide clear pathways for requesting or escalating to human assistance and consider these needs and preferences when allocating that support.
    %%The perceived presence of a human also introduced social expectations that were not always beneficial. Some participants felt more motivated to work harder because another person was helping them, while others worried about making mistakes or asking too many questions. [talk about psychological freedom, impression management, and evaluation apprehension] Prior educational research has found that perceived social threats can contribute to avoidance of academic help-seeking \cite{ryan1997should}. In contrast, these participants described the AI tutor as less judgmental and easier to approach. Similar findings have been observed outside education: Lucas et al. found that participants were more willing to disclose information to a virtual interviewer when they believed it was controlled by a computer rather than a human, partly because the computer framing reduced fears of negative evaluation \cite{lucas2014only}. Together, these findings suggest that human interaction can provide interpersonal connection and accountability while also introducing social pressures, whereas perceived AI interaction may reduce some of those pressures.
    
    %At the same time, our finding that human presence can induce anxiety suggests an opportunity to explore systems that allow students to move easily from AI assistance to a human tutor according to their needs and preferences.

    Finally, participants' skepticism toward some AI-generated social cues should not be interpreted as a reason to remove warmth and encouragement from AI tutoring.  Reactions were heterogeneous: although some participants regarded AI social cues artificial or strange, others found its encouragement helpful, and participants generally evaluated both tutors favorably.  In other words, our findings suggest that AI emulation of human behavior is potentially \textit{less} effective for some students, not that it is \textit{ineffective}.  Future work can explore heterogeneous student preferences for chatbot behavior, the underlying reasons, and effects when these preferences are violated.  This could then inform chatbot designers on how to best personalize AI behaviors to support students, rather than merely reproducing human behaviors.
    
    %Our findings strongly suggest that producing socially supportive language may not be sufficient for an AI tutor to provide the same sense of connection as a human.  

% Let students adjust tone, friendliness, praise, and level of detail (not everyone wants human-like AI interaction)
% Make encouragement meaningful i.e. specific feedback on progress/reasoning rather than generic praise or forced empathy.
% Let students choose how much help they want (small hints vs. detailed explanations). Avoid giving away answers prematurely. -- also in the autonomy lit
% AI should acknowledge its own limitations rather than making students feel responsible for poor communication. -- future work needs to know how much this moves the needle on mental models, because so many other things affect it (other models, the news, and longer term interaction)
% AI for low-pressure practice/questions; humans for personal connection, nuanced discussion, encouragement.  -- Well-supported by the literature as well, but we have good triangulation here
% Preserve student choice i.e. different students prefer different types of support; don't force everyone into the same AI tutoring experience. --needs more context for this advice

\section{Limitations}
    %What direction is the bias, and which situations are most generalizable?
    One limitation is that we did not include a manipulation check assessing whether participants believed the assigned tutor identities. Several participants remarked that the human tutor's behavior resembled AI, whereas they seldom described the opposite.  If some participants suspected that the ostensibly human tutor was AI, this ambiguity may have lessened differences between conditions.  In a more naturalistic setting, tutor identity would likely be reinforced by additional cues, which could strengthen identity-based differences on our measured outcomes.  However, this would also bring instructional quality, availability, and other factors into play.
    
    We also caution against overgeneralizing this study in three major aspects.  First, our findings are limited to brief, initial interactions with a purported AI tutor and an unfamiliar human tutor.  Over a longer period, students could become less willing to rely on an AI tutor after observing errors, as suggested by research on algorithm aversion \citep{algoAversion,dzindolet2003trust}.  Repeated interaction with a human tutor could also allow greater familiarity and interpersonal connections to develop a stronger sense of relatedness in students, while companionship with AI might develop differently.  Second, these findings are limited to tutoring in subjects in which student responses can be objectively tested.  For open-ended questions in other subjects such as history and literature, subjective feedback could complicate trust dynamics between tutors and students.
    %Lastly, these findings are limited to one approach to tutor behavior.  We took an autotomy-supportive stance, where we carefully provided hints to preserve opportunities for students to solve problems by themselves. Different kinds of tutor behavior (e.g. a more casual stance) might mediate the effects we saw.

% Different school subjects and different kinds of tutor behavior (e.g. a more casual stance) might mediate the effects we saw.

    Lastly, our participants' responses were likely shaped by their existing expectations of human and AI behavior.  These expectations may change as students gain experience with increasingly capable AI systems. Consequently, the particular behaviors that participants regarded as characteristically human, artificial, or supportive may not generalize unchanged to future systems or student populations.  However, our broader finding is not tied to any single set of expectations: perceived tutor identity provided a lens through which participants interpreted similar instructional and social behaviors and decided how to communicate.

    %Lastly, our findings are mediated by student expectations of human and AI behaviors.  As students' experiences with AI and their expectations of it develop, so might their views.

    %There are a couple of factors that we did not assess or measure in this study due to time limits or technical difficulties in our current setting. We did not assess students' prior exposure to and their perceived competence of AI models. Past interactions with highly error-prone vs. highly capable models might shift how students perceive the AI tutor in this experiment. As students' expectations of AI develop, so might their views. We also did not assess whether students believed that the human was human and the AI was AI. If students thought that the human was AI, then we would have *underestimated* the effect size. If students thought that the AI was human, then ??? As for student motivation, which is theorized to be the downstream outcome of satisfying psychological needs, we did not make an objective assessment. Although several students mentioned changes in their motivations, these self reports could be biased by social desirability and limited self-awareness. A more reliable motivation assessment would require behavioral measures that include time spent on the task and voluntary practices.
%But if anything, students thought the human was AI.  This would mean that we *underestimate* the effect size.

\bibliography{refs}

\appendix

\section{Survey items}
\label{appendix:survey_items}

Participants indicated 7-point Likert agreement for all items.  For \textbf{perceived autonomy support}, items were as follows:

\begin{enumerate}
    \item I feel that the tutor provided me choices and options.
    \item I feel understood by the tutor.
    \item The tutor conveyed confidence in my ability to do well.
    \item The tutor encouraged me to ask questions.
    \item The tutor listened to how I would like to do things.
    \item The tutor tried to understand how I saw things before suggesting a new way to do things.
\end{enumerate}

\noindent For \textbf{relatedness need satisfaction}:
\begin{enumerate}
    \item I really liked the tutor.
    \item I got along with the tutor.
    \item The tutor cares about me.
    \item The tutor does not seem to like me much (reverse coded).
\end{enumerate}

\noindent For \textbf{competence need satisfaction}:
\begin{enumerate}
    \item I felt a sense of accomplishment from completing the problems in this session.
    \item I felt very competent in my ability to complete the problems in this session.
\end{enumerate}

\noindent For \textbf{overall enjoyment}:
\begin{enumerate}
    \item I was satisfied with the tutor.
    \item I enjoyed working with the tutor.
\end{enumerate}

\end{document}